\documentclass{article}%
\usepackage{amsfonts}
\usepackage{amsmath}
\usepackage{amssymb}
\usepackage{graphicx}%
\begin{document}

\title{Inducing charges and currents from extra dimensions}
\author{M. A. S. Cruz$^{a}$, F. Dahia$^{b}$ and C. Romero$^{a}$\\$^{a}$Departamento de F\'{\i}sica, Universidade Federal da Para\'{\i}ba, \\Caixa Postal 5008, 58059-979 Jo\~{a}o Pessoa, Pb, Brazil\\E-mail: cromero@fisica.ufpb.br \\$^{b}$Departamento de F\'{\i}sica, Universidade Federal de Campina\\Grande, 58109-970, Campina Grande, Pb, Brazil\\E-mail: fdahia@df.ufcg.edu.br}
\maketitle

\begin{abstract}
In a particular variant of Kaluza-Klein theory, the so-called induced-matter
theory (IMT), it is shown that any configuration of matter may be
geometrically induced from a five-dimensional vacuum space. By using a similar
approach we show that any distribution of charges and currents may also be
induced from a five-dimensional space. Whereas in the case of IMT the geometry
is Riemannian and the fundamental equations are the five-dimensional Einstein
equations in vacuum, here we consider a Minkowskian geometry and
five-dimensional Maxwell equations in vacuum.

\end{abstract}

\section{Introduction}

Certainly it is not easy to assess the wealthy and considerable influence that
the ideas put forward by Kaluza-Klein theory \cite{Kaluza,Klein} in the early
twentieth \ century have had on the development of modern theories
of\ physics, particularly in their quest for unification of the fundamental
interactions of matter. In Kaluza-Klein theory one would say that a certain
kind of unification of gravity and electromagnetism is achieved. This
accomplishment, however, has a price: we need to introduce an extra (as yet
unseen) dimension to ordinary spacetime. On the other hand, this artifice
allowed T. Kaluza and O. Klein to devise a mechanism through which an extra
dimensionality of space combines with curvature in such a way that
electromagnetic phenomena may be looked upon as a manifestation of pure geometry.

The idea that we might live in a five-dimensional (5D) space \ instead of the
traditional four-dimensional (4D)\ picture and that some observable physical
effects may be attributed to the existence of an extra dimension was seminal.
It endures until these days and lies at the core of many
currents\ research\ which, one would say, constitute the 'main stream' of
theoretical particle physics and cosmology \cite{Collins}. Inspired by the
five-dimensional Kaluza-Klein theory there appeared, in the sixties and
seventies, a number of higher-dimensional theories such as eleven-dimensional
supergravity and ten-dimensional superstrings, all of them aiming at a general
scheme of unification \cite{Collins}. More recently, another
higher-dimensional model, the so-called \textit{braneworld }scenario has
emerged, according to which our 4D spacetime is viewed as a hypersurface
(\textit{the brane}) isometrically embedded\textit{ }in an five-dimensional
Einstein space (\textit{the bulk}) \cite{Randall,Randall2}.

The original version of Kaluza-Klein theory assumes, as a postulate, that the
fifth dimension is compact. Recently, however, a non-compactified approach to
Kaluza-Klein gravity, known as Induced-Matter theory (IMT) has been proposed
by Wesson and colaborators \cite{Wesson,Overduin,Wesson2}. The basic principle
of the IMT approach is that all classical physical quantities, such as matter
density and pressure, are susceptible of a geometrical interpretation.
Moreover, it is asserted that only one extra dimension is sufficient to
explain all the phenomenological properties of matter in a geometrical way. In
more specific terms, IMT proposes that the classical energy-momentum tensor,
which appears on the right-hand side of the four-dimensional Einstein
equations could be, in principle, generated by pure geometrical means. Much in
the same spirit of Kaluza-Klein theory, Wessons's proposal also assumes that
the fundamental five-dimensional space in which our usual spacetime is
embedded, should be a solution of the five-dimensional vacuum Einstein equations%

\[
R_{ab}=0.
\]

Surely along the history of physics there have been many attempts to formulate
a unified field theory which would not resort to extra dimensions. In fact,
recourse to an unobserved extra dimension was basically the cause of
Einstein's reluctance to accept the plausibility of Kaluza-Klein theory
\cite{Pais}. Alternative ways to geometrize the electromagnetic field in the
usual four-dimensional spacetime have been proposed since the early days of
general relativity. Among these is, for instance, the work of Weyl, who in
1918 proposed a kind of generalization of Riemannian geometry \cite{Weyl}, not
to mention the attempts made by Einstein, Eddington, Schr\"{o}dinger and many
others \cite{history}.

An aspect of Kaluza-Klein theory that perhaps has gone unnoticed concerns the
role played by the geometry upon which the\ underlying fundamental
higher-dimensional theory is based. In the case of Kaluza-Klein this
fundamental theory is general relativity, recasted in a five-dimensional
Riemannian space. Here lies the mysterious power of the Kaluza-Klein program.
Indeed, were not for the richness Riemannian geometry has comparatively
to\ flat geometric background, there would hardly be enough available degrees
of freedom capable of accomodating in its geometrical structure
non-gravitational fields such as the electromagnetic field. This understanding
leads to the epistemological question of what really is the role and the power
of the extra dimensional hypothesis, taken by itself, in our quest for
formulating unifying theories. In other words, to what extent is the apparent
success of Kaluza-Klein theory (or its extension to more general gauge fields
\cite{ModernKaluza})\ a consequence of the extra-dimensional hypothesis alone?

In this paper we shall try to give a partial answer to the above question\ by
examining\ a theory in which the original Kaluza-Klein scheme has been
modified. That is, instead of assuming the (vacuum) Einstein field equations
as the fundamental equations, that are capable of generate electrodynamics in
4D, or matter (in the Induced-Matter theory approach), we promote the (vacuum)
Maxwell equations\ in 5D to the level of fundamental equations and explore
their potentiality to generate new physics in 4D. For simplicity we shall work
in a five-dimensional flat space background.\ It is a curious question whether
the Kaluza-Klein mechanism or the Wesson's procedure for 'generating' matter
from the extra dimension still works. In other words, we would like to answer
the question: When the fundamental 5D equations are Maxwell equations, instead
of Einstein equations, is it possible to 'induce' electric charge and currents
in 4D from pure vacuum in 5D?\ \ \ \ \ \ \ \ \ \ \ \ \ \ \ \ \ \ \ \ \ \ \ \ \ \ \ \ \ \ \ \ \ \ \ \ \ \ \ \ \ \ \ \ \ \ \ \ \ \ \ \ \ \ \ \ \ \ \ \ \ \ \ \ \ \ \ \ \ \ \ \ \ \ \ \ \ \ \ \ \ \ \ \ \ \ \ \ \ \ \ \ \ \ \ \ \ \ \ \ \ \ \ \ \ \ \ \ \ \ \ \ \ \ \ \ \ \ \ \ \ \ \ \ \ \ \ \ \ \ \ \ \ \ \ \ \ \ \ \ \ \ \ \ \ \ \ \ \ \ \ \ \ \ \ \ \ \ \ \ \ \ \ \ \ \ \ \ \ \ \ \ \ \ \ \ \ \ \ \ \ \ \ \ \ \ \ \ \ \ \ \ \ \ \ \ \ \ \ \ \ \ \ \ \ \ \ \ \ \ \ \ \ \ \ \ \ \ \ \ \ \ \ \ \ \ \ \ \ \ \ \ \ \ \ \ \ \ \ \ \ \ \ \ \ \ \ \ \ \ \ \ \ \ \ \ \ \ \ \ \ \ \ \ \ \ \ \ \ \ \ \ \ \ \ \ \ \ \ \ \ \ \ \ \ \ \ \ \ \ \ \ \ \ \ \ \ \ \ \ \ \ \ \ \ \ \ \ \ \ \ \ \ \ \ \ \ \ \ \ \ \ \ \ \ \ \ \ \ \ \ \ \ \ \ \ \ \ \ \ \ \ \ \ 

\section{The five-dimensional field equations}

We start by assuming that our fundamental space is \ a flat five-dimensional
Lorentzian manifold $M^{5}$ with metric $n_{ab}$ $=$ $diag(+----)$
\footnote{Throughout Latin indices take value in the range (0,1,...4) while
Greek indices run from (0,1,2,3). The operation of raising and lowering
indices is done with the help of the 5D and 4D metrics ($n_{ab,}$ $n_{\mu\nu}%
$) and their inverses ($n^{ab},$ $n^{\mu\nu}$). We shall also denote the fifth
coordinate $y^{4}$ by $l$ and the first four coordinates $y^{\mu}$\ (the
"spacetime" \textit{ }coordinates) by $x^{\mu}$, that is, $y^{a}=(x^{\mu},l)$,
with $\mu=0,1,...,3.)$}. We also take the view, according to modern embedding
theories, that $M^{5}$ is foliated by a family of hypersurfaces $\left\{
\Sigma\right\}  $ defined by $l=const$\ and that our ordinary spacetime
consists of one of these hypersurfaces $\Sigma_{0}$ $(l=0)$, which is then
embedded in $M^{5}$ \footnote{In order to simplify the model we shall neglect
gravity, hence $\Sigma_{0}$ will be identified to 4D Minkowski spacetime}. In
the fundamental space $M^{5}$ we shall postulate the existence of an
'electromagnetic' field, which is described by a \ 5-dimensional potential
vector defined on $M^{5}$ and\ denoted by $A^{a}=(A^{\mu},\Phi).$ \ We then
define the 5D electromagnetic field $\ $tensor $F_{ab}=\partial_{a}%
A_{b}-\partial_{b}A_{a}$ and postulate that the dynamics of $A^{a}$ \ in
$M^{5}$ \hspace{0in}is governed by the vacuum field equations%
\begin{equation}
\partial_{a}F^{ab}=0 \label{Maxwell5D}%
\end{equation}
where $F^{ab}\equiv n^{ac}n^{bd}F_{cd}$.

Let us consider the equations (\ref{Maxwell5D}) separately by first taking
$b=\nu$, \ and then putting$\ b=4$. \ We then have
\begin{equation}
\partial_{\mu}\left(  \partial^{\mu}A^{\nu}-\partial^{\nu}A^{\mu}\right)
=\left(  \partial^{\nu}\frac{\partial\Phi}{\partial l}+\frac{\partial
^{2}A^{\nu}}{\partial l^{2}}\right)  \label{Maxwell4D}%
\end{equation}
and
\begin{equation}
\square\Phi=-\frac{\partial}{\partial l}(\partial_{\mu}A^{\mu}),
\label{Constraint}%
\end{equation}
with $\square\equiv\partial_{\mu}\partial^{\mu}$ denoting the usual
d'Alembertian operator of 4-dimensional spacetime \footnote{In this work we
are employing Gaussian units.}. By proceeding in analogy with Kaluza-Klein
theory (or with the IMT approach) we shall consider the above equations taken
at $l=0$ and then regard Eq. (\ref{Maxwell4D}) as the ordinary 4D Maxwell
equations, where the right-hand side has been formally identified with a
4-dimensional current density vector $j^{\mu}$ \cite{Landau}.

The following question now arises: Can we induce any arbitrary
four-dimensional current density $j^{\mu}=j^{\mu}(x)$ in the same way as
one\ can induce any arbitrary energy-momentum tensor in the IMT approach? In
the case of the Induced-Matter approach \ this question was resolved by
translating the problem into a geometrical language and \ employing the
Campbell-Magaard theorem, which in essence asserts that any $n$-dimensional
Riemannian manifold is locally embeddable into a $(n+1)$-dimensional
Ricci-flat manifold \cite{romero,campbell,Magaard}. Here we shall try a more
direct approach by investigating the system of partial differential equations
defined by (\ref{Maxwell4D}) and $\ $(\ref{Constraint}), and look for
solutions $A^{\mu}=A^{\mu}(x,l)$, $\Phi=\Phi(x,l)$ for a given set of
prescribed functions $j^{\mu}=j^{\mu}(x)$. Let us note that since $j^{\mu}$ is
to be interpreted physically as a current density it has to satisfy the
equation of continuity $\partial_{\mu}j^{\mu}(x)=0$.

For simplicity let us first set $\Phi=0$ (we shall see later that this is a
possible choice). Then (\ref{Maxwell4D}) becomes
\begin{equation}
\frac{\partial^{2}A^{\nu}}{\partial l^{2}}=\partial_{\mu}\left(  \partial
^{\mu}A^{\nu}-\partial^{\nu}A^{\mu}\right)  \label{Maxwell2}%
\end{equation}
We note that according to the Cauchy-Kowalevskaya theorem \cite{Cauchy} the
equation (\ref{Maxwell2}) admits an analytical solution in a neighbourhood of
$l=0$.

It turns out that a explicit solution of (\ref{Maxwell2}) may be found in the
form of the power series
\[
A^{\mu}\left(  x,l\right)  =\sum_{n}a_{\left(  n\right)  }^{\mu}\left(
x\right)  l^{n}%
\]
where the coefficients $a_{\left(  n\right)  }^{\mu}$ are calculated by a
process of repeated iteration, as we show below. If (\ref{Maxwell2}) is to
induce the 4D Maxwell equations for an arbitrary prefixed current density
$j^{\mu}=j^{\mu}(x)$ in the hypersurface $l=0$, then we must have
\begin{equation}
\left.  \frac{\partial^{2}A^{\nu}}{\partial l^{2}}\right\vert _{0}=\frac{4\pi
}{c}j^{\nu}\left(  x\right)  \label{2}%
\end{equation}
Now the third derivative may be obtained by taking the derivative of
(\ref{Maxwell2})\ with respect to $l$. In this way we have:%
\[
\left.  \frac{\partial^{3}A^{\nu}}{\partial l^{3}}\right\vert _{0}%
=\partial_{\mu}\left(  \partial^{\mu}\left.  \frac{\partial A^{\nu}}{\partial
l}\right\vert _{0}-\partial^{\nu}\left.  \frac{\partial A^{\mu}}{\partial
l}\right\vert _{0}\right)
\]
Clearly the coefficient $\left.  \frac{\partial A^{\nu}}{\partial
l}\right\vert _{0}$ must be chosen in such a way to satisfy (\ref{Constraint})
. Thus, let us set $\left.  \frac{\partial A^{\nu}}{\partial l}\right\vert
_{0}=\eta^{\nu}$, where $\eta^{\nu}\left(  x\right)  $ is an arbitrary
four-vector with null divergence $\left(  \partial_{\nu}\eta^{\nu}=0\right)
$. With this choice it follows that%
\[
\left.  \frac{\partial^{3}A^{\nu}}{\partial l^{3}}\right\vert _{0}=\square
\eta^{\nu}%
\]
The fourth derivative, on the other hand, gives:%
\[
\left.  \frac{\partial^{4}A^{\nu}}{\partial l^{4}}\right\vert _{0}%
=\partial_{\mu}\left(  \partial^{\mu}\left.  \frac{\partial^{2}A^{\nu}%
}{\partial l^{2}}\right\vert _{0}-\partial^{\nu}\left.  \frac{\partial
^{2}A^{\mu}}{\partial l^{2}}\right\vert _{0}\right)
\]
From (\ref{2}) and recalling that $\partial_{\mu}j^{\mu}=0$, we have%
\[
\left.  \frac{\partial^{4}A^{\nu}}{\partial l^{4}}\right\vert _{0}=\frac{4\pi
}{c}\square j^{\nu}%
\]
Repeating the same argument it is easy to verify that $\left.  \frac
{\partial^{5}A^{\nu}}{\partial l^{5}}\right\vert _{0}=\square\left(
\square\eta^{\nu}\right)  $, $\left.  \frac{\partial^{6}A^{\nu}}{\partial
l^{6}}\right\vert _{0}=\frac{4\pi}{c}\square\left(  \square j^{\nu}\right)  $,
and so on. Therefore, we conclude that $A^{\mu}(x,l)$, solution of
(\ref{Maxwell2}), will be given by%
\begin{equation}
A^{\mu}\left(  x,l\right)  =a^{\mu}\left(  x\right)  +\eta^{\mu}\left(
x\right)  l+\frac{1}{2!}\frac{4\pi}{c}j^{\mu}\left(  x\right)  l^{2}+\frac
{1}{3!}\left(  \square\eta^{\mu}\right)  l^{3}+\frac{1}{4!}\left(  \frac{4\pi
}{c}\square j^{\mu}\right)  l^{4}+... \label{generalsolution}%
\end{equation}
It is important to note that $A^{\mu}(x,l)$ satisfy the equations
(\ref{Maxwell4D}) and (\ref{Constraint}) for all orders of $l.$ Note also that
$a^{\mu}\left(  x\right)  $ is nothing more than the potential induced on the
hypersurface $l=0$ and that $a^{\mu}\left(  x\right)  $ satisfies the 4D
Maxwell equations\ with the source $j^{\mu}(x)$.\ Moreover, we have usual
gauge freedom in the induced 4D electrodynamics since the 4D electromagnetic
field tensor $F^{\mu\nu}(x)=\partial^{\mu}A^{\nu}(x,0)-\partial^{\nu}A^{\mu
}(x,0)$ is clearly invariant under transformations of the type $a^{\mu}\left(
x\right)  \rightarrow a^{\mu}\left(  x\right)  +\partial^{\mu}f\left(
x\right)  .$

If, for some reason, we would like to have an even solution $A^{\mu}\left(
x,l\right)  $ in the $l$ coordinate (in order to have $Z_{2}$ symmetry, for
instance), then we can choose $\eta^{\mu}=0$. In this case the series reduces
to
\[
A^{\mu}\left(  x,l\right)  =a^{\mu}\left(  x\right)  +\frac{4\pi}{c}\sum
_{n=1}\frac{1}{\left(  2n\right)  !}\left(  \square^{n-1}j^{\mu}\right)
l^{2n}%
\]
where the use of the symbol $\square^{n}$ means that the d'Alembertian
operator has been applied $n$ times.

We conclude then that if we are given any analytical four-dimensional vector
function $j^{\mu}=$ $j^{\mu}(x)$, describing a certain physical distribution
of charge and current, then we can produce a five-dimensional potential
$A^{a}=(A^{\mu}(x,l),\Phi(x,l))$, solution of the vacuum Maxwell equations in
5D, with $A^{\mu}(x,l)$ given by (\ref{generalsolution}) and an arbitrary
function $\Phi=\Phi(x,l)$, which for simplicity we have chosen to be$\ \Phi
=0$,\ such that when $A^{a}(x,l)$ is restricted to the hypersurface
$\Sigma_{0}$ $(l=0)$, it generates or induces $j^{\mu}=$ $j^{\mu}(x)$. One
would say then that, similarly to the generation of matter in the
Induced-Matter approach from a vacuum higher-dimensional space, any physical
configuration of charge and current may also be viewed as having a purely
geometric origin due to extra dimensionality of space.

\section{Final remarks}

In this paper we basically have tried to answer the following question: What
is the potentiality of Maxwell equations\ to generate new physics in
four-dimensional spacetime from a five-dimensional vacuum? It is well known
the mechanism by which the Kaluza-Klein theory can generate, via the Einstein
equations, the gravitational and gauge fields in four dimensions from a
higher-dimensional vacuum space. We know that in the Induced-Matter approach,
which also employs the Einstein equations in vacuum, matter can be viewed as
being generated from an extra dimension by purely \ geometrical means. If we
replace the Einstein equations by the vacuum Maxwell equations as the
fundamental equations of our higher-dimensional theory, apparently there are
not enough degrees of freedom in a five-dimensional Minkowskian geometry
capable of generating a metric field describing gravitation plus the
four-dimensional electromagnetic field \footnote{In this respect it should be
mentioned the work of G. Nordstr\"{o}m, which was the first attempt to
formulate a unified theory of gravitation and electromagnetism by postutating
the existence of a fifth dimension. However, in his theory spacetime is flat
and gravitation was described by means of a scalar field. } \ \cite{Nordstrom}%
. Nevertheless we discover quite surprisingly that, analogously to what
happens in the case of the induced-matter approach, although no other field is
generated, charges and currents may also be viewed as induced by the extra dimension.

We would like to mention that when the present article was almost ready we
became aware of a recent paper by Liko, in which the idea of inducing an
effective electromagnetic current \ from higher dimension is also taken up
\cite{Liko}. In this approach the mentioned author considers a 4D
electromagnetic tensor \ $\mathcal{F}_{\mu\nu}$ modified by the presence of an
extra term depending non-linearly on the electromagnetic potentials $A^{\mu}$
which are supposed to be functions of the fifth coordinate, while the 4D
Maxwell equations are derived from the variation of a four-dimensional action
built with $\mathcal{F}_{\mu\nu}$. In this case the induced current does not
follow directly from the 5D Maxwell equations in vacuum.

One basic assumption in Kaluza-Klein theory is that the vacuum is the space
$M^{4}\times S^{1}$, i.e the product of 4D Minkowski spacetime with a circle
of radius $R$. In the case of 4D induced electrodynamics since we do not need
the assumption of compactness of the fifth dimension the vacuum should be
considered as the five-dimensional Minkowski space $M^{5}$. As was pointed out
by Gross and Perry \cite{gross}, both vacua are classically stable, although
it has been argued by Witten that the ground state $M^{4}\times S^{1}$ is
unstable against a process of semiclassical barrier penetration \cite{Witten}.

Finally, the question of inducing electromagnetic charges and currents from a
higher-dimensional vacuum space may be given a more geometrical approach by
reducing it to an initial value problem for the electromagnetic field, much in
the same way as the Induced-Matter theory may be formulated in the light of
the Campbell-Magaard theorem \cite{Dahia}. In a quantum context it is worth
mentioning that the idea of a 4D electromagnetic field embedded in a 5D pure
vacuum has been considered very recently by Raya, Aguilar and Bellini in
connection with inflation and gravitoelectromagnetic effects \cite{Aguilar}.

\section{Acknowledgement}

The authors would like to thank CNPq-FAPESQ (PRONEX) for financial support.
Thanks also go to the referee for useful and relevant comments.

\end{document}